\pdfoutput=1

\documentclass[12pt]{article}

\usepackage{amsfonts}
\usepackage{amsmath}
\usepackage{amssymb}
\usepackage{bigints}
\usepackage{booktabs}
\usepackage[nosort]{cite}
\usepackage{color}
\usepackage{dsfont}
\usepackage{float}
\usepackage{framed}
\usepackage{graphicx}
\usepackage{indentfirst}
\usepackage{mathrsfs}
\usepackage{multirow}
\usepackage{pdflscape}
\usepackage{setspace}
\usepackage{subdepth}
\usepackage{subfig}
\usepackage{titlesec}
\usepackage[dotinlabels]{titletoc}
\usepackage{wrapfig}
\usepackage[all]{xy}
\usepackage{young}
\usepackage[vcentermath]{youngtab}
\usepackage{relsize}
\usepackage{stackengine}
\usepackage{datetime}

\usepackage{hyperref}
\usepackage{physics}
\usepackage[dvipsnames]{xcolor}

\usepackage{mathtools}

\numberwithin{equation}{section}

\usepackage{verbatim}

\newcommand{\be}{\begin{equation}}
\newcommand{\ee}{\end{equation}}

\newcommand{\bml}{\begin{multline}}
\newcommand{\emll}{\end{multline}}

\newcommand{\nn}{\nonumber}

\def\({\left(} \def\){\right)}
\def\[{\left[} \def\]{\right]}

\def\al{\alpha}

\def\mO{\mathcal{O}}

\def\eps{\epsilon}

\def\v{\vec}

\def\mD{\mathcal{D}}

\def\a{\alpha}

\def\g{\gamma}

\def\lam{\lambda}

\newcommand{\G}{\Gamma}

\def\d{\partial}

\def\o{\omega}

\newcommand{\la}{\langle}
\newcommand{\ra}{\rangle}

\newcommand{\bea}{\begin{eqnarray}}
\newcommand{\eea}{\end{eqnarray}}

\usepackage[left=2cm,right=2cm,top=2cm,bottom=2cm]{geometry}
\titleformat{\section}{\large\bfseries}{\thesection.}{4pt}{}
\titlespacing{\section}{0pt}{22pt}{6pt}

\titleformat{\subsection}{\normalfont\bfseries}{\thesubsection.}{4pt}{}
\titlespacing{\subsection}{0pt}{18pt}{6pt}

\titleformat{\subsubsection}{\normalfont\itshape}{\thesubsubsection.}{4pt}{}
\titlespacing{\subsubsection}{0pt}{16pt}{6pt}

\def\ie{\begin{equation}\begin{aligned}}
\def\fe{\end{aligned}\end{equation}}

\def\tilde{\widetilde}

\def\hat{\widehat}

\def\bar{\overline}

\usepackage{tikz}
\tikzset{every picture/.style={line width=0.75pt}} 

\def\d{\partial}

\def\1{{\mathds 1}}

\def\mL{\mathcal{L}}

\def\o{\omega}
\def\v{\vec }

\DeclareFontShape{OT1}{cmr}{mx}{n}%
    {<->cmr10}{}
\newcommand{\mytitlefont}{\fontseries{mx}\selectfont}
\DeclareMathAlphabet{\titlemath}{OT1}{cmr}{mx}{n}

\newcommand{\bi}{\begin{itemize}}
\newcommand{\ei}{\end{itemize}}

\newcommand{\sss}{\subsubsection}

\usepackage{bm}

\def\sss{\subsubsection}

\begin{document}

\begin{titlepage}

\begin{center}

~\\[1cm]

{\fontsize{20pt}{0pt} \mytitlefont Renormalization Group in far-from-equilibrium states}\\[10pt]

~\\[0.2cm]

{\fontsize{14pt}{0pt}Vladimir Rosenhaus{\small $^{1}$} and Michael Smolkin{\small $^{2}$}}

~\\[0.1cm]

\it{$^1$ Initiative for the Theoretical Sciences}\\ \it{ The Graduate Center, CUNY}\\ \it{
 365 Fifth Ave, New York, NY 10016, USA}\\[.5cm]
 
 \it{$^2$ The Racah Institute of Physics}\\ \it{The Hebrew University of Jerusalem} \\ \it{
Jerusalem 91904, Israel}

~\\[0.6cm]

\end{center}

\noindent

We study renormalization group flows in far-from-equilibrium states. The study is made tractable by focusing on states that are spatially homogeneous, time-independent, and scale-invariant. Such states, in which mode $k$ has  occupation numbers $n_k \sim k^{-\gamma}$, are well known in nonlinear physics. RG flow in such states is qualitatively different from that  in the vacuum ---  a positive $\gamma$ decreases the dimension of an operator, turning marginal interactions into relevant interactions.
We compute one-loop beta functions. Depending on the sign of the beta function, backreaction may either cause a minor shift of the state in the IR, or completely change the nature of the state. Focusing on nearly marginal interactions, we construct an analog of the epsilon expansion and IR fixed points, with epsilon now set by the scaling of the interaction rather than the spacetime dimension. In the language of RG flow, critical-balance scaling -- which has applications in fields as varied as astrophysics and ocean waves -- corresponds to the state dynamically adjusting itself along the RG flow until the interaction becomes marginal.

\vfill 

\end{titlepage}

\tableofcontents
~\\

\section{Introduction}
A paradigm of quantum field theory is that one defines the theory in the UV and then seeks to extract the IR behavior. Tree-level renormalization group (RG) dictates which terms appearing in the UV Lagrangian can and cannot affect the IR physics: the relevant operators can, the irrelevant ones cannot. Often, one can have interactions that are marginal, and quantum (loop) corrections are necessary. The interactions in both QED and QCD are marginal but have vastly different behavior in the IR: the interaction in QED decays in the IR, while for QCD it grows.  The growth in the QCD case means the system is very different from a free field theory, serving as a precursor of confinement.

Of course, this analysis and these statements are state-dependent and assume that the state is the vacuum. For instance, QCD in a high temperature thermal state exhibits deconfinement rather than  confinement.  Likewise, while a four fermion interaction term added to QED in the vicinity of the vacuum is irrelevant, in a Fermi liquid state there can be excitations along the Fermi surface with no energy cost, rendering four-fermion interactions marginal \cite{Shankar:1993pf, Polchinski:1992ed}. If the interaction grows in the IR one has BCS superconductivity, much like confinement, instead of gapless free-fermion excitations that would have  otherwise been expected. 

The leading order analysis of the relevance/irrelevance of an operator is no more than dimensional analysis. For instance, for relativistic, massless, $\lam \phi^4$ theory, the strength of the interaction is set by the ratio of the quartic to quadratic terms in the Hamiltonian, $\phi^4/(\d \phi)^2 \sim  k^{-2} \phi^2 \sim k^{D-4}$, where $D$ is the spacetime dimension. This holds in the vicinity of the vacuum; for a general state, in which mode $k$ has  occupation number $n_k$, we replace $\la \phi^2\ra \sim (n_k+\frac{1}{2})/\o_k$, yielding $ (n_k+\frac{1}{2}) k^{D-4}$ for the strength of the nonlinearity. For a general $n_k$ there is not much more one can say. However, suppose $n_k$ scales as a power law, $n_k \sim k^{-\g}$. Then, for large $n_k$, this ratio becomes $k^{D-4 - \g}$. For positive $\g$, the interaction shifts from being marginal in $D=4$ to relevant. 

The goal of this paper is to study renormalization group flows for marginal and nearly marginal operators in states with scale-invariant occupation numbers, $n_k \sim k^{-\g} \gg 1$. We will consider several examples and in each ask the usual question: is the beta function  positive or negative? A negative beta function indicates that the state is significantly modified in the IR.

The scale-invariant, stationary, far-from-equilibrium states, $n_k \sim k^{-\g}$, are well-known in nonlinear physics \cite{Falkovich, Zakharov}, and have been extensively studied analytically, numerically, and experimentally in a diverse set of examples such as: waves in the ocean \cite{hasselmann_1962, Newell, zakharov2018analytic}, vibrations of elastic plates \cite{DURING201742, Hassaini_2019}, Langmuir waves in plasmas \cite{zakharov1972collapse}, Bose-Einstein condensates \cite{OT, Gaz19, zhu2024, FR2}, the early universe \cite{Micha:2004bv, Chatrchyan:2020cxs},   heavy-ion collisions \cite{Berges:2020fwq}, and many others.

In Sec.~\ref{Sec2} we consider the standard relativistic $\lam \phi^4$ field theory, using the Keldysh formalism to compute the beta function in scale-invariant states. In Sec.~\ref{Sec3} we compute the beta function  in far-from-equilibrium states in theories with an arbitrary scale-invariant quartic interaction and dispersion relation. In both Sec.~\ref{Sec2} and Sec.~\ref{Sec3}, the scaling exponents of the interaction, dispersion relation, and the state are such that the interaction is marginal in the chosen state. In Sec.~\ref{Sec4} we consider nearly marginal interactions in states $n_k\sim k^{-\g}$, setting up an analog of the epsilon expansion, commonly employed to study the flow from a free theory to a weakly interacting IR fixed point. We conclude in Sec.~\ref{Sec5}. Appendix~\ref{apA} reviews the construction of the scale-invariant, stationary, far-from-equilibrium states. Appendix~\ref{apB} shows how beta functions can be found by summing the leading log-divergent diagrams.

\section{Relativistic scalar field theory} \label{Sec2}
We start with the standard relativistic, massless, scalar quantum field theory with a quartic interaction, 
\be \label{lamphi4}
\mL = \frac{1}{2} (\d \Phi)^2 - \frac{\lam}{4!} \Phi^4~.
\ee
We will be computing correlation functions in an excited state. Since the path integral naturally computes in-out correlation functions, the Keldysh procedure accommodates in-in correlation functions by taking a time contour that runs forward and then backward, so that we start and end on the same initial state. It is standard to denote the field on the upper branch of the contour (running forwards in time) by $\Phi^+$ and the field on the lower branch by $\Phi^-$, so the Lagrangian becomes $\mL(\Phi^+) - \mL(\Phi^-)$, see e.g., \cite{HuRose}. The Green's function  becomes a two-by-two matrix of Green's functions. After a field rotation, 
\be
\phi = \frac{1}{2}(\Phi^+ + \Phi^-)~, \ \ \ \ \ \eta = \Phi^+- \Phi^-~,
\ee
the Lagrangian becomes,
\be \label{etaphiL}
\mL =\d \eta \d \phi - \frac{\lam}{3!} (\eta \phi^3+  \frac{1}{4}\eta^3 \phi)~,
\ee
and the Keldysh and retarded Green's functions take the form, respectively, 
\be\label{GkR}
G_k^K= \la \phi_k^* \phi_k\ra = 2\pi\delta(k^2)(n_{ \bf k}+\frac{1}{2})~, \ \ \ G^R_k = \la \phi_k^* \eta_k\ra = \frac{i}{k^2 + i \eps k_0}~,
\ee
where we are using relativistic notation, $k^2 = k_0^2 - {\bf k}^2$. The interaction term with one $\eta$ field is referred to as the classical interaction vertex, as it survives in the classical limit, while the interaction term with $\eta^3$ is the quantum interaction vertex. 

We  assume that we are in a state that is time-independent and Gaussian, in which mode $k$ has occupation number $n_{\bf k}\sim {|\bf k|}^{-\g}$, with $n_{k}\gg 1$. When computing the Keldysh Green's function from the free action, it is the boundary conditions at the initial time that establish the $(n_{\bf k}+1/2)$ factor appearing in the expression. From the form of the correlators (\ref{GkR}), we can assign the following dimensions (in momentum space) to $\phi$ and $\eta$, which we denote by $\Delta_{\phi}, \Delta_{\eta}$, respectively:~\footnote{We are dropping the $1/2$ in the Keldysh Green's function, since we are assuming $n_{\bf k}\gg 1$. The appearance of $D$ in the dimension is because of the momentum conserving delta function i.e., what is really meant by the correlator is
$\la \phi_k^* \phi_p\ra = n_{\bf k} \delta(k^2)\delta^{D}(k{-}p)$.}
\be
\ \  \ \ \ \ \ \ \ \ \ \  \ \ \ \ \ \  \ \ \ \ \ \ \  \  \ \Delta_{\phi} = - \frac{\g +D+2}{2}~, \ \ \ \Delta_{\eta}=  \frac{\g -D-2}{2}~, \ \ \  \ \  \ \ \  \ \ \ \ \text{momentum space}~.
\ee
Scaling dimensions are more commonly written in position space, and  are related to the momentum space dimensions through $\Delta\rightarrow D-\Delta$, giving the position space dimensions,
\be \label{26}
\ \  \ \ \ \ \ \ \ \ \ \  \   \  \Delta_{\phi} =  \frac{D-\g-2}{2}~, \ \ \ \Delta_{\eta}= \frac{\g +D-2}{2}~, \ \ \  \ \  \ \ \  \ \ \ \  \text{position space}~.
\ee
As a consistency check, setting $\g = 0$  recovers the standard (vacuum) scaling dimensions, while for general $\g$ we have $\Delta_{\phi} + \Delta_{\eta} = D-2$, which is necessary from (\ref{etaphiL}).

Denoting the scaling dimension of the interaction $\int d^D x\, \eta \phi^3$ by $\mD$ we have, 
\be \label{mD0}
 \mD=3 \Delta_{\phi} + \Delta_{\eta} -D= D-4- \g~.
\ee
Notice that if we are in the vacuum  ($\g = 0$), the interaction is marginal in four dimensions. Likewise, if $\g>0$, the interaction  is relevant in four dimensions. 
Let us therefore suppose that we are in dimensions greater than four, so that $D-4 - \g = 0$ and the interaction is marginal. Notice that (\ref{26}) then gives $\Delta_{\phi}=1$, regardless of the spacetime dimension.

\subsection{One-loop beta function} \label{sec21}
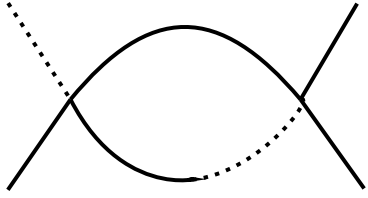
\begin{figure}[t] \centering

\tikzset{every picture/.style={line width=0.75pt}} 

\begin{tikzpicture}[x=0.75pt,y=0.75pt,yscale=-1,xscale=1]

\draw [line width=1.5]    (225.59,138.95) -- (194,184.57) ;
\draw [line width=1.5]    (341.59,138.95) .. controls (300.29,89.91) and (265.49,90.42) .. (225.59,138.95) ;
\draw [line width=1.5]    (284.65,179.69) .. controls (300.29,175.65) and (254.39,194.82) .. (225.59,138.95) ;
\draw [line width=1.5]  [dash pattern={on 1.69pt off 2.76pt}]  (341.59,138.95) .. controls (349.66,135.92) and (321.47,177.16) .. (284.65,179.69) ;
\draw [line width=1.5]    (370.14,90.57) -- (341.59,138.95) ;
\draw [line width=1.5]    (341.59,138.95) -- (373.67,183.88) ;
\draw [line width=1.5]  [dash pattern={on 1.69pt off 2.76pt}]  (194.12,90.33) -- (225.59,138.95) ;

\end{tikzpicture}
\caption{Momenta $q$ inside the loop are confined to a shell $\mu{-}\delta \mu<q<\mu$. Solid lines denote $\phi$, while dotted lines denote $\eta$.} \label{phi4etaphi}
\end{figure}

Focusing on the $\eta \phi^3$ term in the Lagrangian (\ref{etaphiL}), in the case that it is marginal,  we now compute the one-loop beta function to determine whether it is marginally relevant or marginally irrelevant. We do this in the context of Wilsonian RG. 

A correlation function of some operator $\mO$, which is e.g., a product of fields $\phi$ and $\eta$ at some momenta, is given by the path integral,
\be
\la \mO \ra = \int \prod_p d\phi_p d\eta_p\, \mO e^{i S}~.
\ee
It is useful to first perform the path integral over all momenta greater than some  scale $\mu$, where $\mu$ is greater than the scale of any momenta appearing in $\mO$. We denote the resulting action by $S_{\mu}$, 
\be \label{Wilsonian}
\la \mO \ra = \int \prod_{p=0}^{\mu}  d\phi_p d\eta_p\, \mO e^{i S_{\mu}} = \int \prod_{p=0}^{\mu-\delta \mu}  d\phi_p d\eta_p\, \mO e^{i S_{\mu -\delta \mu}}~,
\ee
where in the second equality we  further integrated  over a shell of momenta $\mu{-}\delta \mu<p<\mu$. We let the couplings in the action $S_{\mu}$ be labeled by $\lam(\mu)$. Our goal is to find the behavior of $\lam(\mu)$. 

We now do this explicitly for the $\eta\phi^3$ term in (\ref{etaphiL}), which  in momentum space  takes the form,
\be \label{210}
S_{\mu}= -\frac{1}{3!}\int \prod_{i=1}^4\frac{d^D p_i}{(2\pi)^D}\,\delta(p_1{+}p_2 {-}p_3{ -}p_4) \lam(\mu) \eta_1 \phi_2 \phi_3 \phi_4\, + \ldots 
\ee
where $\phi_i \equiv \phi_{p_i}$ and $\eta_i \equiv \eta_{p_i}$. 
Integrating out a shell $\mu{-}\delta \mu<p<\mu$  gives the relation between $\lam(\mu)$ and $\lam(\mu{-} \delta \mu)$, 
\be \label{lameff0}
\lam(\mu{-}\delta\mu) = \lam(\mu) - \mL~,
\ee
where $\mL$ is the loop integral (see Fig.~\ref{phi4etaphi}),
\be
\!\!\!\!\!\mL =-3 \lam(\mu)^2\! \int_{\mu-\delta \mu}^{\mu} \frac{d^D q}{(2\pi)^D} \frac{(n_{\bf q}+\frac{1}{2})2\pi \delta(q^2)}{(p {-} q)^2 {+} i\eps (p{-}q)^0}{\Bigg|_{p=0}}
= -3 \lam(\mu)^2\! \int_{\mu-\delta \mu}^{\mu} \frac{d^D q}{(2\pi)^D} \frac{(n_{\bf q}{+}\frac{1}{2})2\pi \delta(q^2)}{(q^0{-}i\eps)^2{-}{\bf q}^2}~,
\ee
where in the second equality we took the external momentum $p$ to zero, since by assumption $p \ll \mu$. 
As the figure shows, inside the loop there is one Keldysh propagator and one retarded propagator. 
Factorizing the delta function, 
\be
\delta (q^2) = \frac{1}{2 |{\bf q}|}\(\delta(q^0 - |{\bf q}|) + \delta(q^0 + |{\bf q}|)\) ~,
\ee
we may perform the $q^0$ integral, 
\be
\label{loop}
\mL
=  { 3\over 4} \lam(\mu)^2 \int_{\mu-\delta \mu}^{\mu} \frac{d^d q}{(2\pi)^d} \frac{(n_{\bf q}+\frac{1}{2})}{|{\bf q}|^3}~,
\ee
where $d$ is the spatial dimension, $D=d{+}1$. In our case $n_{\bf q} \sim |{\bf q}|^{-\g}$ where $\g = d{-}3$, making the interaction  marginal. In addition, we are interested in the classical regime, in which $|{\bf q}|$ is sufficiently small so that $n_{\bf q} \gg 1$, allowing us to drop the $1/2$ in the above expression. Hence,
 \be \label{216}
\mL \approx  \frac{3\lam^2}{4} S_d\int_{\mu-\delta \mu}^{\mu}  \frac{dq}{q} \approx  \frac{3\lam^2}{4} S_d\frac{\delta \mu}{\mu}~, \  \ \ \ \ \ \  \ S_d = \frac{2}{(4\pi)^{d/2} \G(d/2)}~.
\ee
We therefore have
\be
\lam(\mu-\delta \mu) = \lam(\mu) - \frac{3 \lam^2}{4} S_d \frac{\delta \mu}{\mu}~,
\ee
which gives the beta function, 
\be \label{beta}
 \mu \frac{d \lambda}{d\mu}   = \frac{3S_d}{4} \lam^2~.
 \ee
 
As a consistency check, to recover the standard vacuum beta function in four dimensions, we set $n_{\bf q}=0$ in (\ref{loop}). This results in (\ref{beta}) with an extra power of $1/2$: $\mu \frac{d \lambda}{d\mu}   = \frac{3S_3}{8} \lam^2$, which is correct since $S_3 = 1/2\pi^2$.

\subsubsection*{Thermal State}
Now, consider the thermal state, 
\be
n_{\bf k} = \frac{1}{e^{|{\bf k}|/T} - 1}~.
\ee
At high energies, $|{\bf k}|/T \gg 1$ and $n_{\bf k}$ decays exponentially: the beta function is simply the vacuum beta function. On the other hand, in the high-temperature limit, $|{\bf k}|/T \ll 1$, the Bose-Einstein distribution reduces to the Rayleigh-Jeans distribution,
\be  \label{RJ}
n_{\bf k} = \frac{T}{|\bf k|}~,
\ee
which is a power-law distribution with $\g =1$. This means that the dimension of the interaction $\mD$ in (\ref{mD0}) is $\mD = D{-}5$. The interaction is marginal in five spacetime dimension, where we can apply the result  (\ref{beta}) for the beta function,
\be \label{thermalbeta}
\mu \frac{d \lambda}{d\mu}   = \frac{3}{32 \pi^2} T\lam^2~,
\ee
 where we  added the normalization factor of $T$ in $n_k$, which was not included in the derivation of (\ref{beta}). 

To be clear: in four spacetime dimensions and at high energies ($\mu \gg T$), the beta function in the thermal state is the standard beta function for $\lam \phi^4$ field theory. At low energies ($\mu \ll T$), the thermal state follows a power-law occupation number (\ref{RJ}). In five spacetime dimensions, the interaction in this state is marginal, with the beta function given by (\ref{thermalbeta}). While the vacuum beta function describes the flow of the coupling due to interactions with quantum vacuum fluctuations, the beta function (\ref{thermalbeta}) describes the running of the coupling due to interactions with thermal bath fluctuations.

Furthermore, the discussion of marginality in this and the next section is a formal idealization introduced to make the one-loop structure explicit. In this sense, we effectively treat the system as being tuned near marginality, even though in physical realizations the system may lie far from this regime.

\section{General Case}\label{Sec3}
In this section, we consider a general complex scalar field with a quartic interaction governed by a Hamiltonian of the form:
 \be
 H = \int\! \frac{d^d k}{(2\pi)^d} \o_k |\Psi_k|^2+ \int \prod_{i=1}^4 \frac{d^d p_i}{(2\pi)^d}(2\pi)^d\delta(\v p_{12;34}) \lam_{p_1p_2p_3p_4} \Psi_{1}^{\dagger} \Psi_{2}^{\dagger} \Psi_{3} \Psi_{4}~,
 \label{Hamiltonian}
 \ee
where $\v k$ is the spatial momentum, $d$ is the spatial dimension, and we introduced the notation $\v p_{12;34} \equiv \v p_1{+}\v p_2{-}\v p_3{-}\v p_4$. Unlike the case in the  previous section, here there is no assumption of Lorentz invariance. The interaction and dispersion relation are homogeneous functions of degree $\alpha$ and $\beta$, respectively: $\o_{k}= k^{\alpha}$ and $\lam_{sp_1 sp_2 sp_3 sp_4}=s^\beta \lam_{p_1p_2p_3p_4} $. The Lagrangian for  this theory is given by:
 \be
  L=-i \int\! \frac{d^d k}{(2\pi)^d} \dot \Psi_k^\dagger \Psi_k - H~.
 \ee
 
 The simplest example of such a system is the nonlinear Schr\"odinger equation, which has a  constant interaction, $\lam_{p_1p_2p_3p_4}=\lam$, and a quadratic dispersion relation, $\o_k = k^2$. The results in this section can  be easily generalized to cases in which the interaction term has an unequal number of $\Psi$ and $\Psi^{\dagger}$ terms. This, in turn, can be used to describe the relativistic $\lam \phi^4$ theory from the previous section, by regarding  $\Psi_k$ as the annihilation operator for mode $k$.
  
As in the previous section, we define the fields $\Psi^{\pm}$ on the upper and lower branches of the Keldysh contour and introduce new variables given by their sum and difference, 
 \be
 \psi_k = \frac{1}{\sqrt{2}}\(\Psi_k^+ + \Psi_k^-\)~, \ \ \  \ \ \eta_k =  \frac{1}{\sqrt{2}} \(\Psi_k^+ - \Psi_k^-\)~.
 \ee
 In terms of these new variables, the Lagrangian $L(\Psi^+)- L(\Psi^-)=L_{\text{free}}+L_{\text{int}} $ is given by \cite{HuRose},
  \bea\nn
 \!\!\!\!  \!\!\!    L_{\text{free}} &=&i \int\! \frac{d^dk}{(2\pi)^d}\, \( \eta_k^{\dagger}(\d_t {+} i \o_k) \psi_k + \psi_k^{\dagger}(\d_t {+} i \o_k ) \eta_k
   \) \\
 \!\!\!\!  \!\!\! L_{\text{int}} 
    &=&\!\! - \int \prod_{i=1}^4 \frac{d^d p_i}{(2\pi)^d}\,  (2\pi)^d\delta(\v p_{12;34}) \lambda_{1234} \(\eta_1^{\dagger} \psi_2^{\dagger} \psi_3 \psi_4+
    \psi_1^{\dagger} \eta_2^{\dagger} \eta_3 \eta_4
    + \text{h.c.}\)~.
    \label{Lint}
\eea
The propagators of the various fields are obtained by inverting the differential operator in $L_{\text{free}}$ and imposing the boundary conditions associated with the in-state,
\bea \label{GkGr}
 G^K_{k, \o} = \la \psi _{k, \o}  \psi_{k, \o}^{\dagger}\ra= (2n_k{+}1) 2\pi \delta(\o{-}\o_k)~, \ \ \ \ \ G^A_{k,\o} &=&\la \eta_{k, \o} \psi_{k,\o}^{\dagger}\ra = \frac{i}{\o - \o_k - i \eps}
\nn\\
G^R_{k,\o} &=&\la \psi_{k, \o} \eta_{k,\o}^{\dagger}\ra = \frac{i}{\o - \o_k + i \eps} ~,
 \eea
 where our  Fourier transform convention is $f_{k,\omega}=\int dt e^{i\omega t} f_k(t)$.  As in the previous section, we assume  we are in a regime  in which the occupation numbers are scale-invariant, $n_k \sim k^{-\g}$. Based on the propagators, we assign scaling dimensions to the fields $\psi_k(t)$ and $\eta_k(t)$:
 \be \label{dim}
\Delta_{\psi} = - \frac{\g +d}{2}~, \ \ \ \Delta_{\eta}=  \frac{\g -d}{2}~.
\ee
The scaling dimension of the interaction terms in $\int dt L_{\text{int}}$ that are linear in $\eta$ -- and which are relevant in the classical (large occupation number $n_k\gg1$) limit --  is then given by,
\be \label{mD}
\mD= 3d+\beta + 3\Delta_{\psi} + \Delta_{\eta} -\al = \beta+ d-\g - \al ~,
\ee
where we used that time scales as $1/\o_k \sim k^{-\al}$.

\subsection*{Scale-invariant far-from-equilibrium states}
So far, we have simply assumed that the state has power-law occupation numbers, $n_k \sim k^{-\g}$. Let us now discuss how such a state can be achieved. One possibility is the Rayleigh-Jeans distribution, $n_k = T/\o_k$ --- the Bose-Einstein distribution at high temperature ---  discussed earlier, see (\ref{RJ}). There is, however, a richer and more interesting possibility. 

One may pump energy into the system at some scale $k_{IR}$. Due to  nonlinearity, the energy cascades into the UV, where it is absorbed at some scale $k_{UV}$. This is referred to as a direct cascade. Alternatively, one may pump in the UV, causing  the wave action to cascade into the IR --- an inverse cascade. If one sets up  $k_{UV}$ and $k_{IR}$ such that the interaction is weak at both these scales, then one can analytically derive a stationary state, $n_k \sim k^{-\g}$, where the exponent $\g$ is determined by the parameters of the theory, see Appendix~\ref{apA}. 
 Specifically, for a direct or an  inverse cascade, the scaling exponent $\g$ and the corresponding dimension of the interaction (\ref{mD}) are, respectively, 
\bea \label{311}
\text{ Direct cascade}~, \ \ \ \g &=& d +\frac{2}{3}\beta~, \ \ \ \ \   \ \ \ \ \ \  \ \mD = \frac{\beta}{3} - \al\\ \label{312}
\text{ Inverse cascade}~, \ \ \ \g &=& d +\frac{2}{3}\beta - \frac{\al}{3}~, \ \ \ \ \ \mD = \frac{1}{3}(\beta -2 \al)~.
\eea
In the special case of a marginal interaction, $\mD = 0$, these become, 
\bea \nn
\text{ Direct cascade}~, \ \ \ \g &=& d +2\al~,  \ \ \ \ \ \ \ \mD = 0 ~,\\
\text{ Inverse cascade}~, \ \ \ \g &=& d +\al~\ \ \ \ \ \ \ \  \ \ \mD = 0~. \label{DIC}
\eea

Let us now turn to computing the beta function in the case of marginal interactions. 
As we implement renormalization group flow and integrate out the UV modes, the functional form of the interaction $\lam_{1234}$ will in general change, already at one-loop level. We begin with a special form of $\lam_{1234}$, which is preserved by the one-loop beta function. 
\subsection{Product factorized coupling} \label{sec31}
We take  an interaction $\lam_{1234}$ that is a product of the magnitudes of the momenta \cite{MMT, ZAKHAROV2001573}, 
\be \label{315}
\o_k = k^{\al}~, \ \ \ \ \lam_{1234} = \lam(p_1^2 p_2^2 p_3^2 p_4^2)^{\beta/8}~.
\ee
Concretely, an example of an interaction of this form -- which is marginal for an inverse cascade -- is a Hamiltonian  that is local in position space,\footnote{This Hamiltonian is a toy model, introduced for its pedagogical simplicity in illustrating momentum-dependent running.}
\be
H = \int d^d x \( |\nabla^2 \Psi|^2 +\lam |\nabla^2 \Psi|^4\)~.
\ee
In momentum space this corresponds to the parameters (\ref{315}) with $\al = 4$ and $\beta = 8$. 

Let us now compute the one-loop beta function, keeping general $\al$ and $\beta$. 
As in Sec.~\ref{sec21}, we will look at the term in the  action that is linear in $\eta$, 
\be \label{316}
S_{\mu}= -\int \prod_{i=1}^4 \frac{d^d p_i d\o_i}{(2\pi)^{d+1}}\,  (2\pi)^{d+1}\delta(\v p_{12;34})\delta(\o_{12;34}) \lambda_{1234}(\mu) \big(\eta_1^{\dagger} \psi_2^{\dagger} \psi_3 \psi_4 + \psi_1^{\dagger} \psi_2^{\dagger} \eta_3 \psi_4 \big)
\ee
where $\psi_i \equiv \psi_{k_i, \o_i}$, $\eta_i \equiv \eta_{k_i, \o_i}$, and $\o_{12;34}\equiv\o_1{+}\o_2{-}\o_3{-}\o_4$. 
\begin{figure}[t] \centering
\subfloat[]{ \includegraphics[width=1.5in]{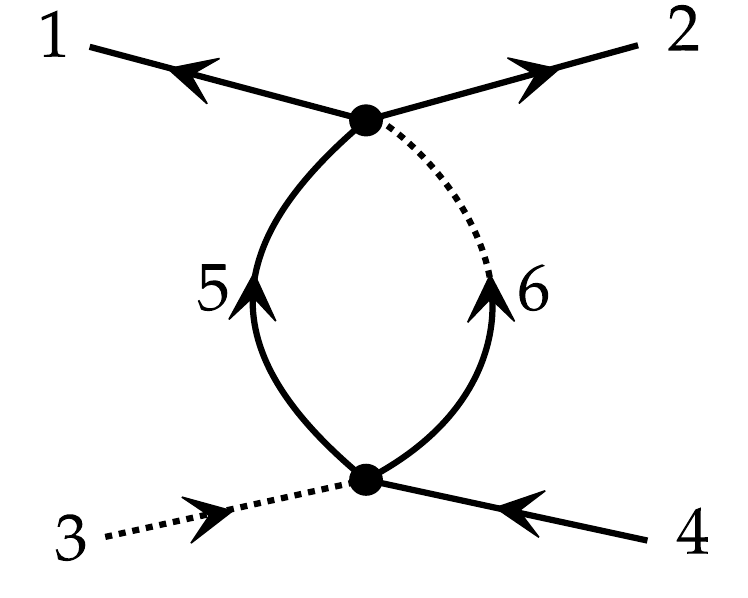} } \ \ \ \ \ \ \ \ \ \ \ \ \ \ 
\subfloat[]{ \includegraphics[width=1.5in]{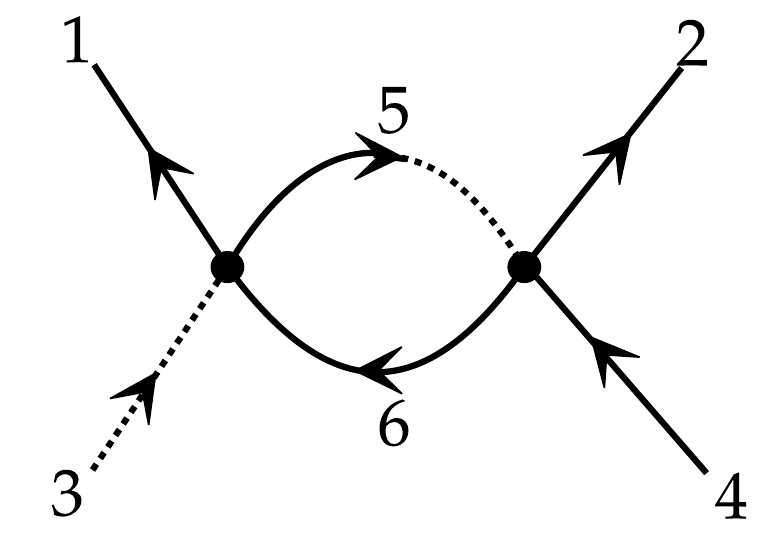} }
\caption{The (a) $s$ channel and (b) $t$ channel one-loop diagrams. The $\psi$ field is represented by a solid line and correspondingly the solid lines are Keldysh propagators, while the $\eta$ field is represented by a dashed line and the solid/dashed lines are advanced propagators, (\ref{GkGr}). The momenta inside the loop are restricted to a shell of radius $\mu$ with thickness $\delta \mu$.  } \label{stchan}
\end{figure}
Integrating out a shell $\mu{-}\delta \mu<p<\mu$  gives the relation between $\lam_{1234}(\mu)$ and $\lam_{1234}(\mu{-} \delta \mu)$, 
\be \label{lameff}
\lam_{1234}(\mu{-}\delta\mu) = \lam_{1234}(\mu) - \mL_s -\mL_t-\mL_u
\ee
where $\mL_s$, $\mL_t$, $\mL_u$ are the $s$, $t$, and $u$ channel  loop integral, respectively, see Fig.~\ref{stchan}. 

Explicitly, the $s$ channel diagram is given by, 
\bea \nn
\mL_s &=&  -2  \int \frac{d\o_5}{2\pi} d\o_6 \int_{\mu{-}\delta \mu}^{\mu}\frac{ d^d p_5}{(2\pi)^d} d^d p_6\, \delta(\o_{12;56}) \delta(\v p_{12;56}) \lam_{1256}{\lam_{5634} \frac{(2n_5{+}1)\delta(\o_5 {-} \o_{p_5})}{\o_6 {-} \o_{p_6} {-}  i\eps}}\\ \label{mLs}
&=&-4 \int_{\mu{-}\delta \mu}^{\mu}\frac{d^d p_5}{(2\pi)^d} d^d p_6  \, \delta(\v p_{12;56}) \lam_{1256}{\lam_{5634} \frac{n_5 {+}\frac{1}{2}}{\o_{12;p_5 p_6} {-}  i\eps}}~,
\eea
where  we made use of the propagators (\ref{GkGr}) and in the second equality we performed the $\o_5$ and $\o_6$ integrals, by using a delta function for one of them and closing the contour to pick up the pole for the other. Since, by assumption, $p_i\ll \mu$, we have that $\v p_6\approx - \v p_5$ and, 
using the form of the interaction (\ref{315}), $\lam_{1234}(\mu) = \lam(\mu) (p_1^2 p_2^2 p_3^2 p_4^2)^{\beta/8}$, we have 
\be \label{319}
\mL_s  \approx 2\lam(\mu)^2 \int_{\mu{-}\delta \mu}^{\mu}\frac{d^d p_5}{(2\pi)^d}\,  p_5^{\beta - \g - \al} = 2\lam(\mu)^2 S_d\int_{\mu{-}\delta \mu}^{\mu} \frac{d p_5}{p_5}\approx2 \lam(\mu)^2 S_d\frac{\delta \mu}{\mu}~,
\ee
where we took the interaction to be marginal, $d {+}\beta {-} \g {-} \al=0$. 

For the $t$-channel diagram, after performing the energy integrals inside the loop, we have the momentum integral, 
\be
\mL_t = -4\int_{\mu{-}\delta \mu}^{\mu}\frac{d^d p_5 }{(2\pi)^d}d^d p_6\,  \delta(\v p_{16;53}) \lam_{1635}{\lam_{2546} \frac{n_6 {-} n_5 }{\o_{1 p_6;3 p_5} {-} i\eps}}~. \label{mLt}
\ee
Again, by assumption, $p_i\ll \mu$, so that $\v p_6\approx \v p_5$. However, we need to be careful, since this approximation causes  the numerator to vanish, $n_6{-}n_5 \approx 0$. We must therefore Taylor expand both the numerator and the denominator. Doing this, and setting $n_5 = p_5^{-\g}$, gives for the numerator, 
\be \label{321}
n_6 - n_5 = \((\v p_5{+}\v p_3{-}\v p_1)^2\)^{-\frac{\g}{2}} - p_5^{-\g}\approx -\g {\v p_5{\cdot}(\v p_3{-}\v p_1)} p_5^{-\g-2}~.
\ee
Expanding the denominator in a similar manner,
\be \label{322}
\o_{ 1 p_6;3 p_5} \approx \begin{cases}\al \v p_5{\cdot}(\v p_3{-}\v p_1)p_5^{\al-2}~,& \ \ \ \ \  \ \ \al>1 \\
\o_1 - \o_3~,& \ \ \ \ \ \ \ \al\leq 1\end{cases}
\ee
As a result, $\mL_t$ has qualitatively different behavior for $\al$ greater than and less than one. For $\al>1$, 
\be
\mL_t =   4\lam(\mu)^2 \frac{\g}{\al} S_d \int_{\mu{-}\delta \mu}^{\mu}\frac{d p_5}{p_5} \approx 4\lam(\mu)^2 \frac{\g}{\al} S_d\frac{\delta \mu}{\mu}~, \ \ \ \ \ \al>1~.
\ee
On the other hand for $\al\leq 1$, $\mL_t$ vanishes, since the integral is odd under $\v p_5 \rightarrow - \v p_5$, 
\be
\mL_t = 4\lam(\mu)^2 {\g\over \o_1{-}\o_3} \int_{\mu{-}\delta \mu}^{\mu} \frac{d^d p_5}{(2\pi)^d p_5}  (\v p_3{-}\v p_1){\cdot} \v p_5\,  p_5^{\alpha-2}=0~, \ \ \ \ \al\leq 1~.
\ee
Finally,  $\mL_u$ is the same as $\mL_t$. Combining all three channels, in total we get that the beta function is,
 \bea \nn
\ \ \ \ \ \ \   \ \ \ \ \ \ \ \ \ \  \ \mu \frac{d \lambda}{d\mu}  &=& 2S_d \lam^2 ~, \ \ \  \ \ \ \ \ \ \ \  \ \ \ \ \ \ \ \ \al\leq 1 \\
\mu \frac{d \lambda}{d\mu}  &=& 2 S_d \lam^2 ( 1 { +} 4 \frac{\g}{\al})~, \ \ \ \ \ \ \  \ \ \al> 1~. \label{betaC}
 \eea
 
If the coupling is positive, the positivity of the beta function means that the coupling  decreases in the IR --- the interaction is marginally irrelevant. A decreasing coupling means an increasing $n_k$, see Appendix~\ref{apA}: the one-loop corrections make $n_k$ steeper than the background state $n_k\sim k^{-\gamma}$. 
For a negative coupling, the positivity of the beta function means that the coupling increases in the IR: the interaction is marginally relevant. An increasing coupling means a decreasing $n_k$: the one-loop corrections make $n_k$ less steep than the background state $n_k\sim k^{-\gamma}$. At  the scale at which the coupling becomes order-one, we lose perturbative control and can no longer trust that the state has occupation numbers $n_k$. Finally, we note that one typically  fixes the prefactor in front of the power law in $n_k$ by imposing that the flux $Q$ (\ref{PQ}) is constant. This gives $n_k \propto Q^{1/3} k^{-\g}$, which adds a factor of $Q^{1/3}$ to the right-hand side of the beta function (\ref{betaC}). 

To recover the zero-temperature beta function, in (\ref{mLs}) we set $n_5$ to zero, giving $1/2$ of the answer in (\ref{319}), while (\ref{mLt}) vanishes. This is consistent with what one obtains when computing directly at zero temperature, without the Keldysh formalism: the Feynman propagator is $\la T \Psi_k^{\dagger} (t_2) \Psi_k(t_1) \ra = \theta(t_{12}) e^{i \o_k t_{12}}$, and only the $s$-channel diagram is nonvanishing, since in the $t$-channel diagram there is a $\theta(t_{12})\theta(t_{21}) = 0$.

\subsubsection*{Effective action}
As one integrates out modes, lowering the scale $\mu$, the action $S_{\mu}$ will contain two classes of other terms, in addition to the interaction  of the form (\ref{316}) that one starts with in the UV. The first are terms which, in position space, have higher powers of derivatives or, equivalently, in momentum space are of the form (\ref{316}) multiplied by powers of $p_i$. These are simple to obtain: in computing, for example, $\mL_s$ in (\ref{mLs}), we neglected the $p_i$ under the assumption that $p_i \ll \mu$. One could have been more systematic,  perturbatively expanding the integrand in powers of $p_i/\mu$, which would yield  the higher derivative terms. Since these terms are higher derivative, they are irrelevant and not important in the IR. 

\begin{figure}[h] \centering
 \includegraphics[width=1.5in]{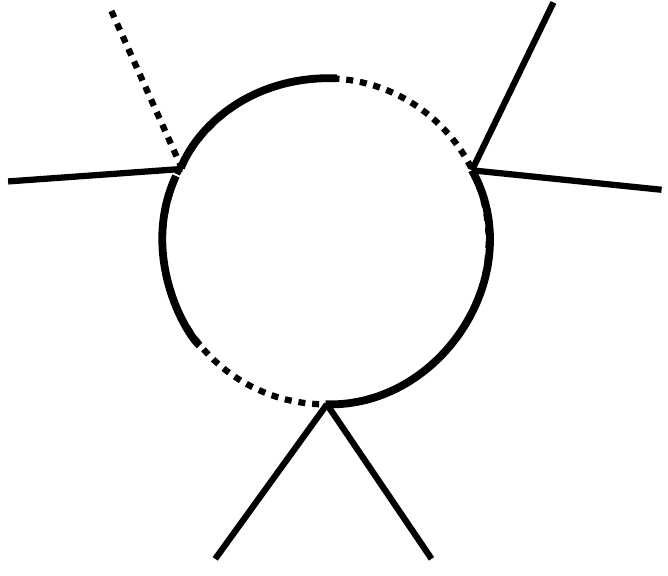}
\caption{Sextic interaction terms generated under RG flow.} \label{sexticFig}
\end{figure}

The effective action $S_{\mu}$ should contain, in addition, terms that are higher order in the number of fields. For instance, the quartic interaction $\lam \eta \psi^3$ in (\ref{316}) will generate an interaction of the form $\lam^3 \eta \psi^5$, see Fig.~\ref{sexticFig}. Ordinarily, interactions with more fields are manifestly irrelevant, since each field comes with a positive dimension. However, in this case, $d{+}\Delta_{\psi}$ (\ref{dim}) is negative. More precisely, the loop diagram generates a sextic interaction term, 
\be \label{sextic}
 \int \prod_{i=1}^6 \frac{d^d p_i d\o_i}{(2\pi)^{d+1}}\,  (2\pi)^{d+1}\delta(\v p_{123;456})\delta(\o_{123;456}) \lam(\mu)^3(p_1 p_2 p_3 p_4 p_5 p_6)^{\beta/4} \mu^{\frac{\beta}{2}-\al}\frac{\delta \mu}{\mu} \big(\eta_1^{\dagger} \psi_2^{\dagger} \psi_3^{\dagger} \psi_4 \psi_5 \psi_6 + \text{h.c.})
 \ee
 where we used that the loop contains one Keldysh propagator and two advanced propagators, leading to the integral
 \be
 \int_{\mu{-}\delta \mu}^{\mu} d^d q \frac{n_q}{\o_q^2} (q^{\beta/2})^3 \sim   \int_{\mu{-}\delta \mu}^{\mu} \frac{d q}{q} q^{d {-}\g{-}2\al {+}\frac{3}{2}\beta} \sim \mu^{\frac{\beta}{2}-\al} \frac{\delta \mu}{\mu}~,
 \ee
 where in the last equality we made use of the quartic interaction being marginal, $\mD = 0$ (\ref{mD}). 
 
 For an inverse cascade, $\mD=0$ requires $\beta = 2\al$ (\ref{312}), so the power of $\mu$ vanishes. Therefore, the generated sextic term is marginal, as opposed to irrelevant as is usually the case in  vacuum RG flow. Nevertheless, this  is sufficient for it to remain small relative to the quartic term (\ref{316}), by nature  of it coming with a $\lam(\mu)^3$. For a direct cascade, on the other hand, $\mD = 0$ requires $\beta = 3\al$, and the sextic interaction (\ref{sextic}) is relevant: deep in the IR our perturbative analysis breaks down, regardless of the sign of the beta function. This is consistent with the picture that RG flow is naturally in the same direction as an inverse cascade (pumping in the UV, dissipation in the IR) and in the opposite direction from  a direct cascade (pumping in the IR, dissipation in the UV)\cite{gawedzki1997inverse}, see also \cite{PhysRevA.25.3281, eyink1994analogies, Canet_2022, Verma:2025rst, PhysRevE.58.1823, goldenfeld2023}.

\subsection{General interaction}
Let us repeat the calculation in the previous section for the one-loop renormalization of the quartic interaction, but now with a general interaction $\lam_{1234}$. 

The flow of the interaction $\lam_{1234}(\mu)$ is again given by (\ref{lameff}), with $\mL_s$ and $\mL_t$ given by (\ref{mLs}) and (\ref{mLt}), respectively. We will obviously be unable to evaluate the angular part of the loop integral, as it depends on the functional form of the interaction. We can, however, evaluate the radial part of the integral since, by assumption, the RG scale $\mu$ is much larger than the momenta  $p_i$. To do this, we need the asymptotic scaling of the interaction in this limit. In particular, we introduce the scaling exponents $\beta_1$ and $\beta_2$, defined as, 
\bea 
&\lim_{p_1,p_3\ll p_2,p_4}\lam_{1234}= (p_2p_4)^{\frac{\beta_1}{2}}(p_1p_3)^{\frac{\beta-\beta_1}{2}} \tilde \lam_{1234}~\label{beta_1}\\
&\lim_{p_1,p_2\ll p_3,p_4}\lam_{1234}= (p_3p_4)^{\frac{\beta_2}{2}}(p_1p_2)^{\frac{\beta-\beta_2}{2}} \bar \lam_{1234}~,  \label{beta_11}
\eea
where $ \tilde \lam_{1234}$ depends only on the ratio of the magnitudes $p_2/p_4$ and $p_3/p_1$, as well as the angles of the $\v p_i$, while $ \bar \lam_{1234}$ depends solely on the ratio of the magnitudes  $p_1/p_2$ and $p_3/p_4$, as well as the angles. 

The momentum integral for  $\mL_s$ (\ref{mLs}) has $\v p_5 \approx - \v p_6 \approx \mu \hat \Omega$, allowing us to replace the couplings,
\be \label{329}
\lam_{1256} \lam_{5634}\rightarrow (p_1 p_2 p_3 p_4)^{\frac{\beta {-}\beta_2}{2}} p_5^{2\beta_2} \bar \lam_{1256}\bar \lam_{5634}\Big|_{\v p_5 = -\v p_6 = \mu \hat \Omega}~.
\ee
Performing the radial integral thus gives $\mL_s$, 
\be \label{gLs}
\mL_s = 2(p_1 p_2 p_3 p_4)^{\frac{\beta {-}\beta_2}{2}}  \frac{\delta \mu}{\mu} \mu^{2\beta_2 - \beta} \int \frac{d \Omega}{(2\pi)^d}  \bar \lam_{1256}\bar \lam_{5634}\Big|_{\v p_5 = -\v p_6 = \mu \hat \Omega}~.
\ee
As a consistency check, for the coupling in Sec.~\ref{sec31}, $\beta_1 = \beta_2 =\beta/2$, $\bar \lam_{1234} = \tilde \lam_{1234} = \lam$, and the angular integral in (\ref{gLs}) simplifies to $\int \frac{d \Omega}{(2\pi)^d}$, so we recover (\ref{319}). 

Likewise, for the $t$-channel, the momentum integral for  $\mL_t$ (\ref{mLt}) has $\v p_5 \approx \v p_6 \approx \mu \hat \Omega$, allowing us to replace the couplings,
\be \label{331}
\lam_{1635} \lam_{2546}\rightarrow (p_1 p_2 p_3 p_4)^{\frac{\beta {-}\beta_1}{2}} p_5^{2\beta_1} \tilde \lam_{1635} \tilde \lam_{2546}\Big|_{\v p_5 = \v p_6 = \mu \hat \Omega}~.
\ee
Performing the radial integral then gives, 
\be \label{gLt}
\mL_t = \begin{cases} \displaystyle 4\frac{\g}{\al}(p_1 p_2 p_3 p_4)^{\frac{\beta {-}\beta_1}{2}}  \frac{\delta \mu}{\mu} \mu^{2\beta_1 - \beta} \int \frac{d \Omega}{(2\pi)^d}  \tilde \lam_{1635} \tilde \lam_{2546}\Big|_{\v p_5 = \v p_6 = \mu \hat \Omega}~, \ \ \ \  \ \ \ \ \ \  \ \ \ \ \ \ \ \ \ \ \  \ \  \  \ \ \ \ \al>1~,\\[15pt]
\displaystyle  4 {\g\over \o_1{-}\o_3} (p_1 p_2 p_3 p_4)^{\frac{\beta {-}\beta_1}{2}}  \frac{\delta \mu}{\mu} \mu^{2\beta_1 - \beta{+}\al{-}1} \int \frac{d \Omega}{(2\pi)^d} \hat\Omega{\cdot}(\v p_3{-}\v p_1)  \tilde \lam_{1635} \tilde \lam_{2546}\Big|_{\v p_5 = \v p_6 = \mu \hat \Omega}~, \ \  \   \al\leq1~,
\end{cases}
\ee
where we made use of (\ref{321}) and (\ref{322}). Compared to the $\al>1$ case, the $\al\leq 1$ case 
has additional factors of $\v p_3{-}\v p_1$ in the numerator and $\o_1{-}\o_3$ in the denominator, which dimensionally are compensated by an extra $\mu^{\al{-}1}$. If $\beta_1=\beta_2$, then for $\al \leq 1$ we see that $\mL_t$ is negligible compared to $\mL_s$, because of the extra $\mu$ suppression. 
Finally, the $u$-channel contribution is obtained by exchanging $3$ and $4$ in $\mL_t$:
\be \label{gLu}
\mL_u = \begin{cases} 
\displaystyle 4\frac{\g}{\al}(p_1 p_2 p_3 p_4)^{\frac{\beta {-}\beta_1}{2}}  \frac{\delta \mu}{\mu} \mu^{2\beta_1 - \beta} \int \frac{d \Omega}{(2\pi)^d}  \tilde \lam_{1645} \tilde \lam_{2536}\Big|_{\v p_5 = \v p_6 = \mu \hat \Omega}~, \ \ \ \  \ \ \ \  \ \ \ \  \ \ \  \ \ \ \  \  \ \ \ \  \ \ \ \al>1~, \\[15pt]
\displaystyle 4 {\g\over \o_1{-}\o_4} (p_1 p_2 p_3 p_4)^{\frac{\beta {-}\beta_1}{2}}  \frac{\delta \mu}{\mu} \mu^{2\beta_1 - \beta{+}\al{-}1} \int \frac{d \Omega}{(2\pi)^d} \hat\Omega{\cdot}(\v p_4{-}\v p_1)  \tilde \lam_{1645} \tilde \lam_{2536}\Big|_{\v p_5 = \v p_6 = \mu \hat \Omega}~ \ \ \al\leq1~.
\end{cases}
\ee

The relation between $\lam_{1234}(\mu{-}\delta \mu)$ and $\lam_{1234}(\mu)$ is given by (\ref{lameff}), with $\mL_s, \mL_t, \mL_u$ given by (\ref{gLs}), (\ref{gLt}), (\ref{gLu}), respectively, with the couplings on the right-hand side evaluated at scale $\mu$. We see that, in general, $\lam_{1234}(\mu{-}\delta \mu)$ assumes  a different functional form than $\lam_{1234}(\mu)$. 
Explicitly, for $\al>1$, the beta function for the coupling function $\lam_{1234}$ is, 
\bea
\mu \frac{d\lam_{1234}}{d \mu} &=& (p_1 p_2 p_3 p_4)^{\frac{\beta {-}\beta_2}{2}} \mu^{2\beta_2 - \beta} \int \!\!\frac{d \Omega}{(2\pi)^d}  \bar \lam_{1256}\bar \lam_{5634}\Big|_{\v p_5 = -\v p_6 = \mu \hat \Omega}\\
&&+4\frac{\g}{\al}(p_1 p_2 p_3 p_4)^{\frac{\beta {-}\beta_1}{2}} \mu^{2\beta_1 - \beta} \int\!\! \frac{d \Omega}{(2\pi)^d} ( \tilde \lam_{1635} \tilde \lam_{2546}+\tilde \lam_{1645} \tilde \lam_{2536})\Big|_{\v p_5 = \v p_6 = \mu \hat \Omega}~,
\eea
with the appropriate replacement if instead $\al\leq 1$. This equation is reminiscent of the flow of the couplings in RG flow on a Fermi surface \cite{SHANKAR1991}.

\subsection{Example with derivative interactions} \label{Sec33}
Let us now study the beta function in another concrete example, with derivative interactions. The Hamiltonian is taken to be, 
\be
H = \int d^d x \( |\nabla \Psi|^2 +\lam_1( \nabla \Psi^* {\cdot} \nabla \Psi)^2 + \lam_2 |\nabla \Psi{\cdot} \nabla \Psi|^2 \)~.
\ee
In momentum space, the dispersion relation is $\o_k = k^2$ and  the interaction takes the form (\ref{Hamiltonian}) with, 
\be \label{lam335}
\lam_{1234} = \frac{\lam_1}{2}( p_1{\cdot}p_3 p_2{\cdot}p_4 +p_1{\cdot}p_4 p_2{\cdot}p_3) + \lam_2 p_1{\cdot}p_2 p_3{\cdot}p_4~.
\ee
The scaling exponents for the dispersion relation and interaction are $\al = 2$ and $\beta = 4$, respectively. This makes the interaction marginal for the state $n_k\sim k^{- d-2}$, corresponding to an inverse cascade (\ref{DIC}). 

This interaction is chosen to preserve its form under one-loop renormalization. The shift symmetry $\Psi\rightarrow \Psi +\text{const.}$ requires that the effective action contains only terms with derivatives of $\Psi$. As we will see, integrating out modes simply redistributes the derivatives among different  $\Psi$ fields, which corresponds to mixing between the $\lam_1$ and $\lam_2$ coupling constants. 

In evaluating the loop integrals, we will need the large $p_5 \gg p_1, {\ldots}, p_4$ limit of the couplings, 
\bea
\lam_{1256}&\rightarrow& - (\lam_1 p_1{\cdot} p_5\, p_2{\cdot} p_5 + \lam_2\,  p_1{\cdot} p_2\,  p_5^2)~, \ \  \ \ \ \  \ \ \  \ \ \ \ \v p_6 \approx - \v p_5\\
\lam_{1635} &\rightarrow & \frac{1}{2} \(\lam_1 p_1{\cdot} p_3\, p_5^2 +(\lam_1 {+} 2\lam_2) p_1{\cdot}p_5\, p_3 {\cdot}p_5\)~, \ \ \ \v p_6 \approx  \v p_5~,
\eea
which will appear in the $s$-channel and $t$-channel loop integrals, respectively. The form of $\mL_s$ from (\ref{gLs}) is then, 
\bea
\mL_s &=&  2\frac{\delta \mu}{\mu} \int\! \frac{d \Omega}{(2\pi)^d}  \frac{\lam_{1256} \lam_{5634}}{p_5^4}\Big|_{\v p_5 = -\v p_6 = \mu \hat \Omega} \\
&=&  2 \frac{\delta \mu}{\mu} \int\!\! \frac{d \Omega}{(2\pi)^d} (\lam_1\,  p_1{\cdot} \hat{\Omega} \, p_2{\cdot} \hat{\Omega} + \lam_2\, p_1{\cdot} p_2 )(\lam_1\,  p_3{\cdot} \hat{\Omega}\, p_4{\cdot} \hat{\Omega} + \lam_2\, p_3{\cdot} p_4 )~,
\eea
while the $t$-channel loop integral $\mL_t$,  given by (\ref{gLt}) with $\al=2$, is 
\bea
\!\!\!\!\!\!\!\!\!\!\!\!\!\!\!\mL_t &=& 4\frac{\g}{2}  \frac{\delta \mu}{\mu} \int\! \frac{d \Omega}{(2\pi)^d}  \frac{\lam_{1635} \lam_{2546}}{p_5^4}\Big|_{\v p_5 =\v p_6 = \mu \hat \Omega} \\
&=&   \frac{\g}{2}\frac{\delta \mu}{\mu} \int\!\! \frac{d \Omega}{(2\pi)^d} \Big(\lam_1\, p_1{\cdot} p_3  +(\lam_1 {+} 2\lam_2) p_1{\cdot}\hat{\Omega}\, p_3 {\cdot}\hat{\Omega}\Big)\Big(\lam_1\, p_2{\cdot} p_4  +(\lam_1 {+} 2\lam_2)\, p_2{\cdot}\hat{\Omega}\, p_4 {\cdot}\hat{\Omega}\Big)~. \ \ \ \ \ \ 
\eea
Making use of the $d$-dimensional angular integrals
\be
\int \frac{d\Omega}{(2\pi)^d} =S_d~, \ \  \int \frac{d\Omega}{(2\pi)^d} \hat{\Omega}_i\hat{\Omega}_j  =S_d\frac{\delta_{ij}}{d} ~, \ 
\int \frac{d\Omega}{(2\pi)^d}   \hat{\Omega}_i\hat{\Omega}_j  \hat{\Omega}_k\hat{\Omega}_l  =S_d \frac{\delta_{ij}\delta_{kl} +\delta_{i k} \delta_{j l} + \delta_{i l}\delta_{j k}}{d(d+2)}~,
\ee
we obtain
\bea
\mL_s &=&\frac{2S_d}{d(d{+}2)} \( ( \lam_1^2+ \lam_2 (d{+}2)(2\lam_1 {+}d \lam_2))    p_1{\cdot}p_2 p_3{\cdot}p_4 +\lam_1^2 (p_1{\cdot}p_3 p_2{\cdot}p_4+p_1{\cdot}p_4 p_2{\cdot}p_3)   \) \frac{\delta \mu}{\mu}\\ \nn
\mL_t &=&\frac{\g}{2}\frac{S_d }{d(d{+}2)} \Big( \big((d^2{+}4d{+}5)\lam_1^2 + 4(d{+}3) \lam_1\lam_2 + 4\lam_2^2\big)    p_1{\cdot}p_3 p_2{\cdot}p_4   \\
&& \ \ \ \ \ \ \ \ \ \  \ \ \ \ \ \ \ + (\lam_1{+} 2\lam_2)^2 ( p_1{\cdot}p_2 p_3{\cdot}p_4 +p_1{\cdot}p_4 p_2{\cdot}p_3) \Big) \frac{\delta \mu}{\mu}~.
\eea
The loop integral in the $u$ channel is simply the $t$-channel result with $3$ and $4$ exchanged. 
  \begin{figure}[t] \centering
 \includegraphics[width=2.5in]{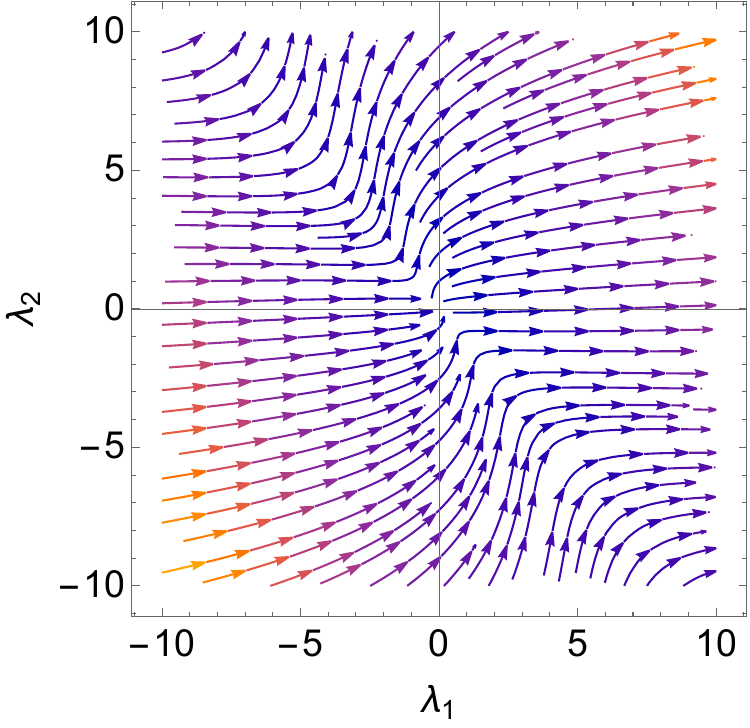} 
\caption{The flow lines for the beta function (\ref{beta2}).} \label{fig:beta2}
\end{figure}
Using (\ref{lameff}) and (\ref{lam335}) we therefore find that the two beta functions for the two couplings $\lam_1, \lam_2$ are, 
  \bea \nn
\mu \frac{d \lam_1}{d\mu} &=&S_d\frac{\g}{d(d{+}2)}\((d^2{+}4d{+}\frac{4}{\g}{+}6) \lam_1^2 + 4(d{+}4) \lam_1 \lam_2 + 8 \lam_2^2  \)  \\  \label{beta2}
\mu \frac{d \lam_2}{d\mu} &=&2S_d \frac{1}{d(d{+}2)} \(  (1{+}\frac{\g}{4})\lam_1^2 + 2(d{+}2{+}\frac{\g}{2}) \lam_1 \lam_2 + (d^2{+}2d {+}\g)\lam_2^2 \)~.
  \eea
  
Using that for an inverse cascade $\g = d{+}2$ (\ref{DIC}), we can make a plot of the beta function in $d=3$, see Fig.~\ref{fig:beta2}. RG flow, moving from the UV into the IR, runs in the  direction opposite to the arrows. Starting with both $\lam_1, \lam_2$ being positive, we see that the couplings flow to zero, similar to the case of one positive coupling in the previous section. On the other hand, starting with one positive and one negative coupling, and following the flow both forwards and backwards, shows that the magnitude of one of the couplings becomes large in both the deep UV and deep IR. In other words, we lose perturbative control.  Unlike what occurred in the single coupling case (\ref{beta}), the background state $n_k$ is valid neither at  very small $k$ nor very large $k$; it is only valid in the vicinity of the starting scale $\mu_0$.

  \section{Nearly marginal interactions} \label{Sec4}
  
Here we turn to interactions that are nearly marginal: the dimension of the interaction (\ref{mD}) is taken to be $\mD = -\eps$, with $\eps \ll 1$. The previous calculations of the one-loop beta function are modified in a simple way,  picking up an extra factor of $\mu^{-\eps}$. Concretely, for couplings of the form $\lam_{1234}(\mu) = \lam(\mu) (p_1^2 p_2^2 p_3^2 p_4^2)^{\beta/8}$ discussed in Sec.~\ref{sec31}, the $s$-channel diagram (\ref{319}) becomes, 
\be
\mL_s  \approx 2\lam(\mu)^2 \int_{\mu{-}\delta \mu}^{\mu}\frac{d^d p_5}{(2\pi)^d}\,  p_5^{\beta - \g - \al}  \approx 2 \lam(\mu)^2 S_d\frac{\delta \mu}{\mu} \mu^{-\eps}~. 
\ee
Likewise, the beta function (\ref{betaC}) is replaced by, 
\be
  \ \ \ \ \ \ \     \ \ \ \ \ \ \ \ \ \   \ \ \ \ \ \ \   \ \ \ \ \ \ \   \ \ \ \     \ \ \ \ \ \ \     \ \ \ \ \ \ \  \mu^{1+\eps} \frac{d \lam}{d\mu} = \kappa \lam^2~, \ \ \ \ \ \  \ \ \ \ \ \  \ \ \ \ \ \ \kappa = 2 S_d \begin{cases} 1~, \ \ \ \ \   \ \ \ \ \al\leq 1\\ 1{+}4\frac{\g}{\al}~, \ \ \ \al>1~,\end{cases}
  \ee
  where we defined $\kappa$ for notional simplicity. 
  Defining a dimensionless coupling $g(\mu) = \lam(\mu) \mu^{-\eps}$ puts the beta function in the more familiar form, 
 \be
  \mu \frac{d g}{d\mu} =- \eps g + \kappa g^2~.
  \ee
  Subsequently integrating the coupling yields,
  \be \label{44}
\lam(\mu) = \frac{\lam(\mu_0)}{1 + \frac{\kappa\lam(\mu_0)}{\eps}(\mu^{-\eps} - \mu_0^{-\eps})}~.
\ee
We see that $\lam$ is given by $\lam(\mu_0)$ at $\mu = \mu_0$, whereas as $\mu\rightarrow 0$ the coupling behaves as, 
\be \label{45}
\lam(\mu\rightarrow0) \rightarrow \frac{\eps}{\kappa} \mu^{\eps}~.
  \ee
  
  The interaction starts off  scale-invariant, with scaling exponent $\beta$ at scale $\mu_0$. It is then scale-dependent at intermediate scales, and becomes scale-invariant again in the IR, now with scaling exponent $\beta+ \eps$. The power of the epsilon expansion is that the coupling remains small throughout. How does the flow of the coupling  influence the state? As reviewed in Appendix~\ref{apA}, (\ref{KZe}), the scaling exponent of the state is determined in part by $\beta$: $\g = d.{+}2\beta/3{-} \a/3$ for an inverse cascade. Therefore, a state that scales as $n_k \sim k^{-\g}$ at the UV scale $\mu_0$ has scaling \cite{Gurarie95}
  \be \label{IR}
  n_k \sim k^{-\g-\frac{2}{3}\eps}
  \ee
  in the IR, as a result of $\beta\rightarrow  \beta{+} \eps$ in the IR. 
  
This conclusion requires that the initial coupling is positive, $\lam(\mu_0)>0$, as otherwise, at low enough $\mu$, $\lam(\mu)$ in (\ref{44}) becomes large and we lose perturbative control. Likewise, while the IR scaling exponent (\ref{IR}) was derived for a specific functional form of the interaction --- products of magnitudes of momenta to a power --- the result is more general. As long as the coupling remains small as one flows into the IR, $\lam(\mu_0)$ will drop out of $\lam(\mu)$ as $\mu\rightarrow 0$ and, on dimensional grounds, this must be replaced by $\mu^{\eps}$, as we saw in (\ref{45}). 
  
    \section{Discussion} \label{Sec5}
  
  The  beta function in vacuum quantum field theory encodes the scale dependence of the coupling as a result of the interaction between a  particle and quantum vacuum fluctuations, causing screening or anti-screening. In a far-from-equilibrium state, the beta function reflects the interaction between the particle and the quantum statistical fluctuations in the background state. We  computed the beta function for several classes of theories in states with scale-invariant occupation numbers $n_k$, starting with a relativistic scalar in Sec.~\ref{Sec2} and then considering more general quartic interactions with arbitrary power law dispersion relations in Sec.~\ref{Sec3}. The flow of the effective coupling with scale  backreacts on the state, causing $n_k$ to become stepper or less steep in the IR, depending on whether the coupling decays or grows in the IR, respectively. 
    
  A frequent question in quantum field theory is the endpoint of RG flow, with options including: a  mass gap, a massless particle, or scale invariance. We may likewise ask: starting with the scale invariant state $n_k \sim k^{-\g}$ in the weakly-coupled UV, what is the behavior of $n_k$ in the IR? In Sec.~\ref{Sec4} we constructed an example of a nearly marginal interaction, with dimension $\epsilon$, that induces a transition from one scale-invariant state in the UV to another in the IR, with  scaling exponents differing by an order $\epsilon$ amount. This can be regarded as analogous to RG flow from a UV fixed point to an IR fixed point within the epsilon expansion. 
  
In real physical systems (e.g., capillary waves or Langmuir-acoustic wave interactions), the parameter $\epsilon$ that measures deviation from the critical balance between dispersion and nonlinearity is not always small. While it is formally introduced as an expansion parameter, it reflects how far the system lies from the critical scaling regime associated with a fixed point. Thus, $\epsilon$ is not freely tunable; rather, it quantifies the system's distance from criticality. However, one can proceed in the spirit of Wilson-Fisher by introducing a formal small parameter controlling this distance, computing perturbative corrections (e.g., to the energy-spectrum exponent), and then extrapolating these results to physical values. This procedure is heuristic in such systems: unlike in standard RG settings, there is no general guarantee that the resulting series can be reliably continued to finite $\epsilon$, and its validity must be assessed a posteriori (e.g., by comparison with numerics or experiments).
  
  Recent work \cite{FR2} addressed a similar problem as the one in this paper: Do one-loop corrections decrease or increase the strength of the interaction and make $n_k$ steeper or less steep, in the context of the nonlinear Schr\"odinger equation?  Explicitly computing the one-loop corrections to the four-point function showed that $n_k$ becomes steeper in the defocusing case (positive coupling) and less steep in the focusing case (negative coupling). The interaction in this case was relevant, with dimension $\mD = -2/3$. In contrast, the interactions studied in this paper were either marginal or nearly marginal, rendering the one-loop integrals  dominated by the UV and simple to evaluate. 
  
 There are several promising future directions: i) The scaling exponent $\g$ in $n_k$ has been measured in a range of experimental setups, both in weakly coupled and strongly coupled regimes, see e.g., \cite{PhysRevFluids.8.014804, FalconMordant, exp2, exp3, PhysRevE.105.L063101, PhysRevLett.133.207201, PhysRevLett.132.224001}. Can the results of Sec.~\ref{Sec3} --- with $\epsilon$ taken to be of order-one --- be matched to these exponents? ii) The exponent $\g$  lowers the dimension of the interaction, making it possible to have marginal interactions with a significant number of derivatives. A four derivative interaction was studied in Sec.~\ref{Sec33}. More derivatives increase the number of marginal operators, which will mix under RG flow, creating a potentially rich set of possibilities, similar to RG flow for theories with multiple fields \cite{Osborn:2017ucf,Osborn:2020cnf,Rychkov:2018vya,Hogervorst:2020gtc,Herzog:2024phd}. iii) While we focused on a single field, many physical applications involve two fields, one of which is usually integrated out, making the remaining interaction appear non-analytic. In the RG framework it would be more appropriate to keep both fields.

In nonlinear systems, it is common to introduce a dimensionless nonlinearity parameter,  heuristically  defined as  the ratio of the interaction energy to the kinetic energy in a state with occupation numbers $n_k\sim k^{-\g}$ \cite{Falkovich, Nazarenko, FR1}. For the Hamiltonian (\ref{Hamiltonian}) the dimensionless parameter is $\eps_k \sim \lam_{k k k k} n_k k^d/\o_k$, involving the ratio of the quartic to quadratic terms with the replacement $|\Psi_k|^2 \sim n_k$. For the scaling $\lam_{k k k k } \sim k^{\beta}$, $\o_k\sim k^{\al}$, this gives $\eps_k\sim k^{\beta{+}d{-}\g{-}\al}$.  This work has made explicit that, in the language of quantum field theory, the dimensionless nonlinearity parameter $\eps_k$ is simply the dimension of the interaction operator in the state with occupation numbers $n_k$. Indeed, the dimension of the interaction (\ref{mD}) is the power of $k$ appearing in $\eps_k$.

For weakly interacting theories, the occupation numbers $n_k$ for a stationary far-from-equilibrium state can be found analytically, as reviewed in Appendix~\ref{apA}. At strong nonlinearity there is of course no clear answer for $n_k$; a widely discussed --- and sometimes empirically successful --- scenario is that of critical balance \cite{Goldreich, Phil, Newell, NS}. Critical-balance scaling postulates that once the nonlinearity parameter becomes of order unity, it stops growing: the system attains ``critical balance'' between the interaction energy and the kinetic energy. With the recognition that the power of $\eps_k$ is the dimension of the interaction, we can restate this in the language of RG flow: critical balance scaling conjectures that in either the UV or the IR, the exponent $\g$ in $n_k\sim k^{-\g}$ takes such a form so that the interaction becomes marginal, i.e. $\g = \beta{+}d{-}\al$. 
This paper has discussed the state dependence of the dimensions of operators. The possibility that the state dynamically changes in order to make interactions marginal is fascinating and worthy of further study. 

\sss*{Acknowledgments} 
We thank Greg Eyink, Grisha Falkovich, Xu-Yao Hu, and Daniel Schubring  for helpful discussions.  This work is supported in part by NSF grant 2209116 and by BSF grant 2022113.

\appendix
\section{Stationary far-from-equilibrium states} \label{apA}
\subsection{Cascade states}

Throughout the main body of the text we have taken it as a given that the state with occupation numbers for mode $k$, $n_k \sim k^{-\g}$ --- with a specific $\g$ --- is a stationary state. Here, we review why this is so. 

The rate of change of $n_k$ for a weakly interacting classical Hamiltonian with a quartic interaction (\ref{Hamiltonian}),  assuming spatial homogeneity, is governed by the wave kinetic equation, 
\be \label{WKE}
    \pdv{n_1}{t} = 16\pi\int \frac{d^d k_2}{(2\pi)^d} \frac{d^d k_3}{(2\pi)^d} \frac{d^d k_4}{(2\pi)^d} \lambda_{1234}^2  n_1 n_2 n_3 n_4\Big( \frac{1}{n_1} {+} \frac{1}{n_2}{-}\frac{1}{n_3} {-} \frac{1}{n_4} \Big)\delta(\o_{k_1 k_2; k_3k_4})\, \delta(\v k_{12;34})~.
    \ee
  This  equation is, in fact, familiar: it is simply the $n_k\gg 1$ limit of the quantum Boltzmann equation for bosons, with the scattering amplitude set equal to its tree level value of $\lam_{1234}$, see e.g., \cite{Kolb}. 
  
There are four distinct stationary solutions of the kinetic equation. Two are trivial: $n_k$ that is constant, and the Rayleigh-Jeans distribution $n_k \sim 1/\omega_k$ (the high temperature limit of the Bose-Einstein distribution). If the interaction and frequency are scale invariant, $\lam_{1234} \sim k^{\beta}$ (meaning, $\lam_{1234}$ is some function of momenta such that under a rescaling of all momenta by a factor of $s$, $\lam_{1234}$ picks up a factor of $s^{\beta}$) and $\o_k\sim k^{\al}$, then there are two more stationary solutions: 
\be \label{KZe}
n_k\sim k^{-\g}~, \ \ \ \ \ \text{where}~ \ \ \  \g =  \begin{cases} d + \frac{2}{3}\beta~, \ \ \ \ \ \  \ \  \  \text{direct cascade}\\
d+\frac{2}{3}\beta - \frac{\al}{3} \ \ \ \ \ \text{inverse cascade}~.\end{cases}
\ee
These are the Kolmogorov-Zakharov (KZ) solutions \cite{Falkovich}; we will give a heuristic derivation of $\g$ in the next subsection. To be precise, if one inserts either of these $n_k$ in (\ref{KZe}) into the kinetic equation (\ref{WKE}), the right-hand side vanishes for all finite $k_1\neq 0$ (provided the integrals converge). Stationary solutions correspond to a constant flux in $|\v k|$ space. Since the $d$-dimensional sphere shrinks to a point at $k=0$, it is not possible to maintain finite flux at $k=0$. Physically, the KZ solution is not supposed to be valid at all $k$ --- to achieve such a solution one always has a source which pumps in energy and a sink which absorbs energy. The KZ solution occurs for wavenumbers that are far separated from both the source and sink. 

The two solutions given in (\ref{KZe}) correspond to the flux of two conserved quantities: energy and total occupation number (wave action). One may compute the sign of the flux, and on the basis of this establish that the energy flux gives a direct cascade (from large to small $k$) whereas the number flux gives an inverse cascade (from small to large $k$). We note that the KZ solutions do not exist in all cases: in addition to requiring that the interaction be weak, the integrals in the kinetic equation must actually converge on the KZ solution. Moreover, the wave must be dispersive: the dispersionless case $\o_{k} = k$ discussed in Sec.~\ref{Sec2}  is outside the regime of perturbation theory \cite{Shi_2016}.

The kinetic equation is a statistical equation,  relying on some sort of averaging. Observationally, one is averaging over space, or over time (for instance, one measures the height of the ocean at some time and does a Fourier decomposition to determine $n_k$, and then repeats at multiple later times) or experimentally, if one is creating the waves in a tank of water, one repeats the experiment many times. Theoretically, the simplest averaging to do is to add a small amount of Gaussian random forcing and dissipation at each wave number, thereby washing out initial conditions and forcing the system into some nontrivial state, and then sending the variance of the forcing and the dissipation to zero,  while maintaining a finite ratio which is chosen self-consistently to give a stationary state \cite{ZakharovLvov, RS1}. This can perhaps be viewed as imitating the noise inherent to any experiment.~\footnote{One should not confuse these ``auxiliary'' forcing and dissipation with physical forcing and dissipation that are added at small and large $k$, respectively, in order to create the far from equilibrium state.} Another theoretical averaging is to choose the initial state to be a Gaussian probability distribution, with an expectation value for the occupation number of mode $k$ that is $n_k$. The different options for what is averaged over have no impact on the properties of the late-time  state \cite{HuRose}. 

The  kinetic equation written in (\ref{WKE}) is only valid to order $\lam^2$, coming from tree-level two-to-two scattering processes. There are higher order in $\lam$ (loop) corrections, which were computed in \cite{RS1,RS2, RSSS} for the classical wave case and in \cite{HuRose} for the quantum case.
The KZ solution is valid as long the higher-order corrections in the kinetic equation are subdominant. The strength of the nonlinearity --- the ratio of the interacting to free parts of the Hamiltonian evaluated in the state --- is a dimensionless quantity with scaling, $\eps_k\sim k^{\mD}$, where $\mD$ is none other than the dimension of the interaction, (\ref{mD}).  Using (\ref{311}) and (\ref{312}) we see that if $\beta>3\al$ ($\beta>2\al$), then the KZ direct cascade (inverse cascade) is valid for small $k$ and breaks down for large $k$. Likewise  if $\beta<3\al$ ($\beta<2\al$), then the KZ direct cascade (inverse cascade) is valid for large $k$ and breaks down for small $k$. Of course, one is free to choose forcing and dissipation to both lie in the regions of weak nonlinearity, so that the breakdown of KZ is never reached. 

\subsection{$q$-body interactions}
Consider a generalization of the quartic interaction studied in the main body, (\ref{Hamiltonian}), to a $q$-body interaction, 
 \be \label{Hintq}
 H_{int} = \int \prod_{i=1}^q \frac{d^d p_i}{(2\pi)^d}(2\pi)^d\delta(\sum_{i=1}^q \v q_i)\, \lam_{12 \cdots q}\, \Psi_{1}^{\dagger} \cdots  \Psi_{q}~,
 \ee 
 where we are agnostic as to the number of creation versus annihilation operators.  Upon switching to the Keldysh $\eta, \psi$ fields, 
the dimension $\mD$ of  the interaction terms linear in $\eta$ is,
\be
\mD= (q{-}1)d+\beta + (q{-}1)\Delta_{\psi} + \Delta_{\eta} -\al = (q{-}2) \frac{d{-}\g}{2} + \beta -\al ~,
\ee
which generalizes (\ref{mD}). 
The energy flux is denoted by $P$, while the number flux is denoted by $Q$, 
\be \label{PQ}
P(k) = \int_0^k d^d q\, \o_{q} \frac{d n_q}{d t}~, \ \ \ \ Q(k) = -\int_0^k d^d q \frac{d n_q}{d t}~.
\ee
Schematically, 
\be
P \sim \frac{n_k k^d \o_k}{\tau_k}~, \ \ \ \  \ \ \ \  Q \sim \frac{n_k k^d }{\tau_k}~,\ \ \ \ \ 
\ee
where the time scale $\tau_k$ is set by the kinetic equation, which is a generalization of (\ref{WKE}) to $q$-body interactions. Schematically, the kinetic equation then gives, 
\be
\frac{n_k}{\tau_k} \sim \frac{\lam_k^2 (n_k k^d)^{q-1}}{\o_k k^d}~,
\ee
where the factor of $\o_k k^d$ in the denominator comes from the energy conserving delta function and the momentum conserving delta function. 
Inserting this $\tau_k$ into the flux gives, 
\be
P \sim \lam_k^2 (n_k k^d)^{q-1}~, \ \ \ \ \ \ \ \ Q \sim \frac{\lam_k^2 (n_k k^d)^{q-1}}{\o_k}~.
\ee
 Setting $P$ or $Q$ to be constant (which corresponds to a stationary solution), and using $\o_k \sim k^{\al}$, $\lam_k \sim k^{\beta}$, gives the KZ solution,
$n_k \sim k^{-\g}$
with
\bea \label{dimqDI}
\text{ Direct cascade}~, \ \ \ \g &=& d + \frac{2}{q{-}1}\beta~, \ \ \ \ \  \ \ \ \ \ \mD = \frac{\beta}{q{-}1} - \al \\
\text{ Inverse cascade}~, \ \ \ \g &=& d + \frac{2\beta {-} \al}{q{-}1}, \ \ \ \ \  \ \ \ \ \ \mD = \frac{\beta}{q{-}1} -\frac{q}{2(q{-}1)} \al~,
\eea 
where the direct cascade corresponds to nonzero $P$ and vanishing $Q$, while the inverse cascade is for nonzero $Q$ and vanishing $P$. The inverse cascade only exists if the total particle number $\int d^d x|\Psi(x)|^2$ is conserved, which requires that the interaction (\ref{Hintq}) have an equal number of creation and annihilation operators. For $q=4$, these $\g$ reduce to (\ref{311}) and (\ref{312}).
In the special case of marginal interactions, $\mD = 0$, the scaling of the KZ state remains (\ref{DIC}), for any $q$. 

Finally, we note from (\ref{PQ}) combined with (\ref{WKE}) that increasing the coupling $\lam_{1234}$ while maintaining constant flux requires decreasing the occupation numbers $n_k$ --- a result we make use of in the main body to connect how the flow of the coupling with scale influences the state.

  \section{Beta function from summing diagrams} \label{apB}
  In this appendix we give an alternative derivation of the beta function in theories with a marginal interaction, by summing the leading log divergent diagrams. 
  \subsection{Scalar field theory in the vacuum}
\begin{figure}[t] \centering
  \subfloat[]{
 \includegraphics[width=3in]{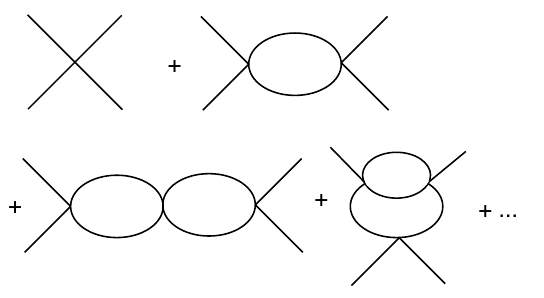} }
   \subfloat[]{
 \includegraphics[width=3.8in]{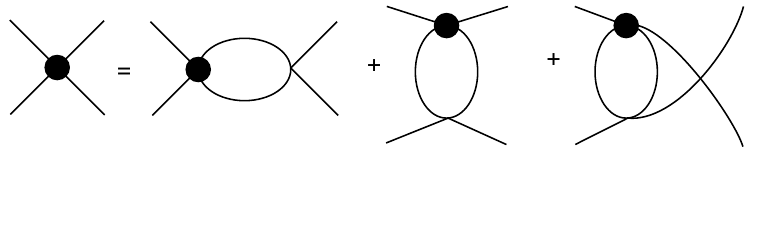}}
\caption{(a) Diagrams up to two loops. (b) A recursion relation to sum all the leading log divergences. } \label{fig5}
\end{figure}

 Let us start with a scalar quantum field theory in four spacetime dimensions with a quartic interaction, (\ref{lamphi4}). At tree level, $\lam(\mu) = \lam_0$, where $\lam_0 \equiv \lam(\mu_0)$. The one loop diagram, shown in Fig.~\ref{fig5}(a), gives the following contribution to $\lam(\mu)$, 
  \be
{-}\frac{3}{2} \lam_0^2 L~, \ \ \ \ \ L =  { \int_\mu^{\mu_0} } \frac{d^4 q}{q^4} = \frac{1}{8\pi^2} { \log {\mu_0\over\mu}}~,
 \ee
 where $\mu$ is the { floating} cutoff.  The factor of $3/2$ arises from: a $1/2$ from the Taylor expansion, a $1/4!$ appearing in the interaction vertex, and a factor of ${4^2\times 3^2\over 2}$ from the number of ways of contracting the legs of the two interaction vertices to form a loop.
 
 The leading divergence of an $n$-loop diagrams scales as $(\log \mu)^n$. A diagram with $n$ loops is given by taking a diagram with $n{-}1$ loops and contracting it with a single vertex, in different possible channels. 
We need to find the sum  of all the leading  $\log$ divergences at each order. Since { divergences are local,} each loop can be contracted to a point, {therefore}, each diagram gives the same contribution. So the problem is a simple counting exercise. Letting $\Lambda_n$ denote the {contribution to $\lam(\mu)$} associated with the sum of the leading log divergences of diagrams { of} order $\lam^n_0$, we have the recursion relation, see Fig.~\ref{fig5}(b),
\be
\Lambda_{n+1} = {-}\frac{3}{2} \Lambda_n \lam_0 L~.
\ee
 The sum over all $n$ is just a geometric series, which gives $\lam(\mu) = \sum_{n=0}^{\infty} \Lambda_n$, 
\be
\lam(\mu) = \frac{\lam_0}{1 { +} \frac{3}{2}\lam_0  L }~.
\ee
Differentiating with respect to $\mu$ gives, 
\be \label{B4}
\mu \frac{d\lam}{d\mu} = \frac{3}{16\pi^2} \lam^2
\ee
which is indeed the beta function for $\lam \phi^4$ theory. 

\subsection{Cascade states}
We may likewise reproduce the results in Sec.~\ref{sec31} for the beta function for interactions of the form (\ref{315}) in  cascade states $n_k \sim k^{-\g}$. We find the recursion relation for the sum of the leading log divergence up to order $n$ is, 
\be 
\Lambda_{n+1} = \begin{cases}~ -2\Lambda_n \lam_0 L~, \ \ \ \  \ \ \  \ \ \ \ \ \al \leq 1~, \ \ \  \ \ \ \ \ L =  \int_{\mu}^{\mu_0} \frac{d^d q}{q^d} = S_d \log \frac{\mu_0}{\mu}~,\\ 
-2(1+ 4 \frac{\g}{\al}) \Lambda_n \lam_0 L~, \ \ \ \ \al >1~.  \label{B5}
\end{cases}
\ee
In the case that $\al\leq 1$, the arrows on every new loop that appears at the next order have to point in the same direction; otherwise, such terms are subleading. In other words, one has to add additional diagrams in the $s$-channel. The only surviving diagrams are then bubble diagrams \cite{Gurarie95}, such as the ones that commonly appear in large $N$ theories
\cite{ Gaz19, berges2002, Walz:2017ffj, bergesRothkopfSchmidt2008, Berges:2010ez, Nowak:2011sk,  B15, RSch}. The combinatorial factor of $2$ in (\ref{B5}) is due to the  choice of pairings between two $a$'s on one vertex and two $a^{\dagger}$'s on the other vertex. For $\al>1$, one must also include the $t$ and $u$-channels, which gives loops with arrows in opposite directions. A combinatorial factor of $4$ is due to the choice  of which $a$ or $a^{\dagger}$ in the first vertex to contract, and there is another combinatorial factor of $2$ from the choice of what it's being contracted  into in the second vertex. The factor of $\g/\al$ comes from the integrand, as discussed in (\ref{321}) and (\ref{322}). Summing all $\Lambda_n$ gives,  
\be
\lam(\mu) =\begin{cases} \displaystyle \frac{\lam_0}{1 +2\lam_0  L } ~, \ \ \ \ \ \ \ \ \   \ \ \ \ \ \ \ \ \ \ \ \ \ \ \al \leq 1~,\\ \displaystyle
\frac{\lam_0}{1 +2(1{+} 4 \frac{\g}{\al})\lam_0  L } ~, \ \ \ \ \  \ \ \ \ \ \ \ \ \ \al >1~.
\end{cases}
\ee
Differentiating reproduces the beta function (\ref{betaC}).

\bibliographystyle{utphys}

\end{document}